\documentclass[twoside,leqno,twocolumn]{article}

\usepackage{times}
\usepackage{amsmath}
\usepackage{amsfonts}
\usepackage{amssymb}
\usepackage{graphicx,psfig,epsfig}
\usepackage{color}
\usepackage{algorithm}
\usepackage{algorithmic}
\usepackage{url}

\newtheorem{theorem}{Theorem}

\newtheorem{proof}{Proof}

\newcommand{\field}[1]{\mathbb{#1}} 

\begin{document}

\title{MACH: Fast Randomized Tensor Decompositions}

\author{Charalampos E. Tsourakakis \thanks{SCS, Carnegie Mellon University} }

\maketitle


\begin{abstract}
Tensors naturally model many real world processes 
which generate multi-aspect data. Such processes appear in 
many different research disciplines, e.g, chemometrics, computer vision, 
psychometrics and neuroimaging analysis. 
Tensor decompositions such as the Tucker decomposition are used to analyze
multi-aspect data and extract latent factors, which capture the multilinear
data structure. Such decompositions are powerful mining tools, for extracting patterns
from large data volumes. However, most frequently used algorithms for such decompositions
involve the computationally expensive Singular Value Decomposition.

In this paper we propose MACH, a new sampling algorithm to compute such decompositions. 
Our method is of significant practical value for tensor streams, such as environmental monitoring systems, 
IP traffic matrices over time, where large amounts of data are accumulated and the analysis 
is computationally intensive but also in ``post-mortem'' data analysis cases where the tensor does not fit in the available memory. 
We provide the theoretical analysis of our proposed method, and verify its efficacy in monitoring system applications.

\end{abstract}

\vspace{2mm} \noindent {\bf Categories and Subject Descriptors:}

\vspace{1mm} \noindent {\bf General Terms:} Algorithms; Experimentation.

\vspace{1mm} \noindent {\bf Keywords: Tensors; Tucker Decompositions; SVD}

\section{Introduction}
\label{sec:intro}
Numerous real-world problems involve multiple aspect data. For example fMRI (functional 
magnetic resonance imaging) scans, one of the most popular neuroimaging techniques, result
in multi-aspect data: voxels $\times$ subjects $\times$ trials $\times$ task conditions $\times$ timeticks. 
Monitoring systems result in three-way data, machine id $\times$ type of measurement $\times$ timeticks.
The machine depending on the setting can be for instance a sensor (sensor networks) or a computer (computer networks).
Large data volumes generated by personalized web search, are frequently modeled as three way tensors, 
i.e., users $\times$ queries $\times$ web pages.


Ignoring the multi-aspect nature of the data by flattening them 
in a two-way matrix and applying an exploratory analysis 
algorithm, e.g., singular value decomposition (SVD) (\cite{Horn:1985:MA}), 
is not optimal and typically hurts significantly the performance (e.g., \cite{Vasilescu02multilinearanalysis}). 
The same holds in the case of applying e.g., SVD on different 2-way 
slices of the tensor as observed by \cite{Kroonenberg}. 
On the contrary, multiway data analysis 
techniques succeed in capturing the multilinear structures in the data, thus achieving
better performance than the aforementioned ideas. 

Tensor decompositions have found the last years many applications in
different scientific disciplines. Indicatively, computer vision and signal processing
(e.g., \cite{Vasilescu02multilinearanalysis,1219330,Sidiropoulos00Blind}), 
neuroscience (e.g., \cite{citeulike:4299268}), 
time series anomaly detection (e.g., \cite{citeulike:3637284}), 
psychometrics (e.g., \cite{tucker3}), chemometrics (e.g., \cite{brobook}),
graph analysis (e.g., \cite{SDM06-LACS,1281266}), data mining (e.g., \cite{hosvd:cubeSVD}). 
Two recent surveys of tensor decompositions and their applications  
are \cite{tamarasurvey},\cite{acar}, with a wealth of references on the topic. 


Two broad families of decompositions are used in the multiway analysis, each with 
its own characteristics: the canonical decomposition (parallel factor analysis), a.k.a. CANDECOMP (PARAFAC) \cite{carroll,harshman}, and the Tucker family of decompositions \cite{tucker3}. In this paper, we focus on the latter. 
The Tucker decomposition can be thought of as the generalization of the Singular Value Decompositions (SVD) to the multiway case. 
Even if there exist algorithms which cast the Tucker decomposition as a nonlinear optimization problem (e.g., \cite{savaslim09}, \cite{SAND2009-0857}),
currently in practice the approach followed is the Alternating Least Squares, which involves the computationally expensive SVD.
To speed up tensor decompositions, randomized algorithms \cite{Drineas05arandomized,DBLP:conf/kdd/MahoneyMD06} have appeared in the recent years. 
This family of randomized algorithms are generalizations of fast low rank approximation methods \cite{1109681, md-cmdid-2009, Drineas04fastmonte}, 
adapted appropriately to the multiway case. 

In this paper we propose a simple randomized algorithm that speedups significantly 
the Tucker decomposition while at the same time results  with guarantees in an 
accurate estimate of the tensor decomposition.
MACH, the proposed method, can be applied both to ``post-mortem'' data analysis 
and to tensor streams to perform data mining tasks
such as network anomaly detection, and in general the set of mining tasks which 
rely on the study of a low rank Tucker approximation. 
MACH is useful when the data does not fit into the available memory and also in 
tensor streams, such as computer monitoring systems, which is also the main motivation behind this work. Specifically,
one of the monitoring systems of Carnegie Mellon University, monitors and uses
data mining techniques to detect failures.
Currently, it monitors over 100 hosts in a prototype data center at CMU. 
It uses the SNMP protocol and it stores the monitoring data in an mySQL database.  Mining anomalies in this system 
is performed using the SPIRIT method and its extension in the multiway case, i.e., the two heads method which uses 
a Tucker decomposition and treats
the time aspect using wavelets \cite{1083674,intemon,citeulike:3637284}. 
Applying the aforementioned methods on large volumes of data is a challenge.

It is worth outlining at this point that in many data mining applications
preserving a constant number of principal components almost the same is of high practical value:
a low rank approximation typically captures a significant proportion of the variance in many real 
world processes and outliers can be detected by examining their position relative to the subspace spanned by the
PCs. 

It is also worth noting that despite many cases
where the formulated tensor is sparse, i.e., few non zero elements as observed in \cite{ICDM08},  there exist 
real world problems where the tensor is dense. As table~\ref{tab:density} shows, for both monitoring system 
we use in the experimental section~\ref{sec:exp}, the resulting tensors are very dense. 
This is the typical case in a monitoring 
system, since at timetick $k$ we receive a measurement of type $j$ for machine $i$, resulting in a non zero in $(i,j,k)$.

\begin{table}[htb] %
\begin{center}
  \begin{tabular}{cc}
      \hline
    \textbf{name} & \textbf{Percentage of non-zeros} 
    \\
    \hline
    {\tt Sensor } &   85 \% \\
    {\tt Network Data} \cite{1316741} &  \\ \hline
    {\tt Computer   } &   81\% \\
    {\tt Network Data (\cite{intemon})} &  \\ 
    \hline
  \end{tabular}
\end{center}
 \caption{Tensors from monitoring system are typically dense.}
\label{tab:density}
\end{table}

The  main contributions of this paper are summarized as in the following: 
\begin{itemize} 
   \item MACH, a randomized algorithm to compute the Tucker decomposition of a tensor ${\cal X}$. 
         MACH is embarrassingly parallel, and adapts easily to tensor streams.
   \item The following theorem, which is our main theoretical result: 
         \begin{theorem}
		 
          Let ${\cal X} \in \field{R}^{I_1 \times I_2 \times \ldots \times I_d}$ a $d$-mode tensor. 
          Let $I_n \geq 76$, $ I_n^2 \leq \prod_{j=1}^d I_j$ for $n=1, \ldots, d$ and 	 
          $b = \max_{i_1,\ldots,i_d} |{\cal X}_{i_1,\ldots,i_d}|$.
     
          For $p \geq  \max_{j} \frac{(8 \sum_{k=1, k \neq j}^d \log{I_k})^4}{(\prod_{k=1,k \neq j}^d I_k)} $ 
          let  ${\cal \hat{X}} \in \field{R}^{I_1 \times I_2 \times \ldots \times I_d}$ be a tensor whose 
          entries are independently distributed as:
          ${\cal \hat{X}}_{i_1,\ldots,i_d} = \frac{{\cal X}_{i_1,\ldots,i_d}}{p}$ with probability $p$, otherwise 0. 
         
		 Let $\breve{{\cal X}}$ be the $(r_1,\ldots,r_d)$-rank approximation of ${\cal \hat{X}}$ given by its HOSVD :
		\begin{equation}
			\breve{{\cal X}}={\cal \hat{X}} \times_1 A^{(1)}A^{(1)T} \times_2  \ldots \times_d A^{(d)}A^{(d)T}
		\end{equation}
		where $A^{(m)}$ is a $I_m \times r_m$ matrix containing the $r_m$ top left singular vectors of the matricization
		of ${\cal \cal{X}}$ along the $m$-th mode. 
	
         Let $X_{(i),r_i}, \hat{X}_{(i),r_i}$ denote the rank $r_i$ approximation of the matricizations $X_{(i)},\hat{X}_{(i)}$ of tensors ${\cal X},
         {\cal \hat{X}}$ along mode $i$ respectively.  
         Then with probability at least  $\prod_{i=1}^d ( 1-exp(-19\sum_{k=1, k \neq i}^d \log{I_k}))$ the following holds:

          \begin{equation}
          || {\cal X} - \breve{{\cal X}}  || \leq t
		  \label{eq:eqt}
          \end{equation}
        
		where $t$ is given by the following equation:

		$t = \min_{i=1\ldots d} t_i$ where 

		$t_i=|| X_{(i)} - X_{(i),r_i} ||+ 4b  (\frac{r_i}{p} \prod_{k=1,k \neq i}^d I_i)^{\frac{1}{2}} +$

		$4 ( ||X_{(i),r_i}|| b )^{\frac{1}{2}}  (\frac{r_i}{p} \prod_{k=1,k \neq i}^d I_i)^{\frac{1}{4}}  + $

        $\sum_{j=1,j\neq i}^d ||\hat{X}_{(j)}- \hat{X}_{(j),r_j} ||$
		
		\label{thrm:thrm1}
		\end{theorem} 
   \item Experiments on monitoring systems, where we demonstrate the success of our proposed algorithm. 
\end{itemize}

The outline of the paper is the following: in Section~\ref{sec:background} we briefly present the necessary theoretical background,
 in Section~\ref{sec:meth} we describe and analyze the proposed method and in Section~\ref{sec:exp} we present the experimental
results. We conclude in Section~\ref{sec:concl}.

\section{Background}
\label{sec:background}	
In this section we briefly present the background behind tensors and low rank approximations. 
Table~\ref{tab:symbol} shows the symbols and the abbreviations we use and their explanation.

\begin{table}[htb]
\begin{center}
\begin{tabular}{|l|l|} \hline
Symbol & Definition and Description \\ \hline \hline
$d$ & number of modes \\ \hline
$I_j$ & dimensionality of \\ 
      &  $j$-th mode \\ \hline 
${\cal X,Y},\ldots \in \field{R}^{I_1 \times \ldots \times I_d}$ & $d$-mode  \\ 
         & tensor (calligraphic) \\ \hline
$ {\cal \hat{X}}$ & tensor obtained upon  \\ 
                    & applying MACH on ${\cal X}$      \\ \hline
$A,U,\ldots \in \field{R}^{m\times n}$ & matrices (upper case)  \\ \hline
$\alpha,\beta,a_{i,j},x_{i_1,\ldots,i_d}$ & scalars (lower case)  \\ \hline
$X_{(i)},\hat{X}_{(i)}$   & matricization of ${\cal X},{\cal \hat{X}}$ \\
            & along mode $i$ \\ \hline
$X_{(i),r_i}, \hat{X}_{(i),r_i}$ &  $r_i$ rank approximation  \\ 
            &  of the matricizations \\
			& $X_{(i)},\hat{X}_{(i)}$ \\ \hline
$\times_n$ & mode-$n$ product  \\  \hline
HOOI & Higher Order Orthogonal \\ 
     & Iteration \cite{354405}  \\  \hline
HOSVD & Higher Order Singular \\
      & Value Decomposition \cite{citeulike:4308452}  \\ \hline
\end{tabular}
\caption{Symbols}
\label{tab:symbol}
\end{center}
\end{table}

\subsection{Tensors}

\begin{figure}[h]
\begin{tabular}{c}
\psfig{figure=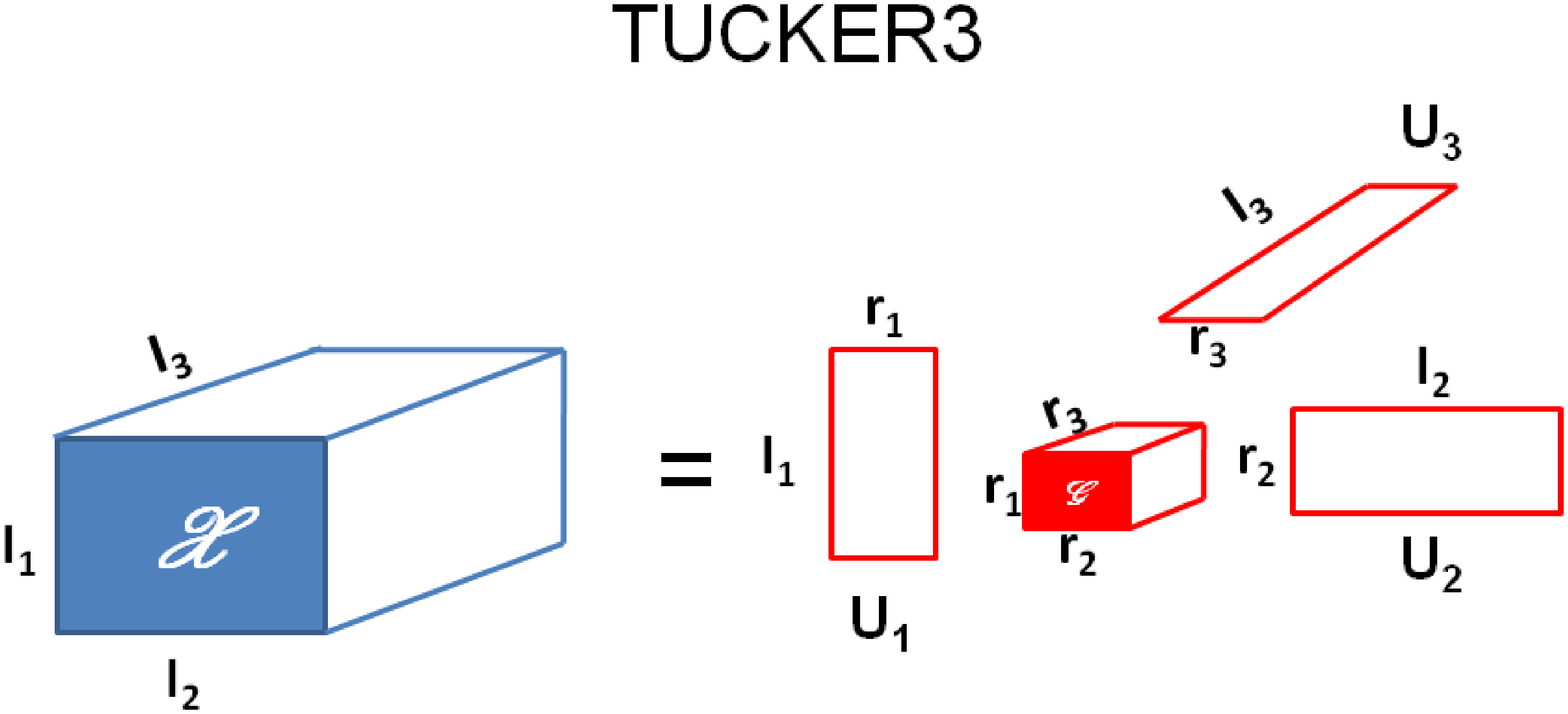,width=0.45\textwidth} \\
\psfig{figure=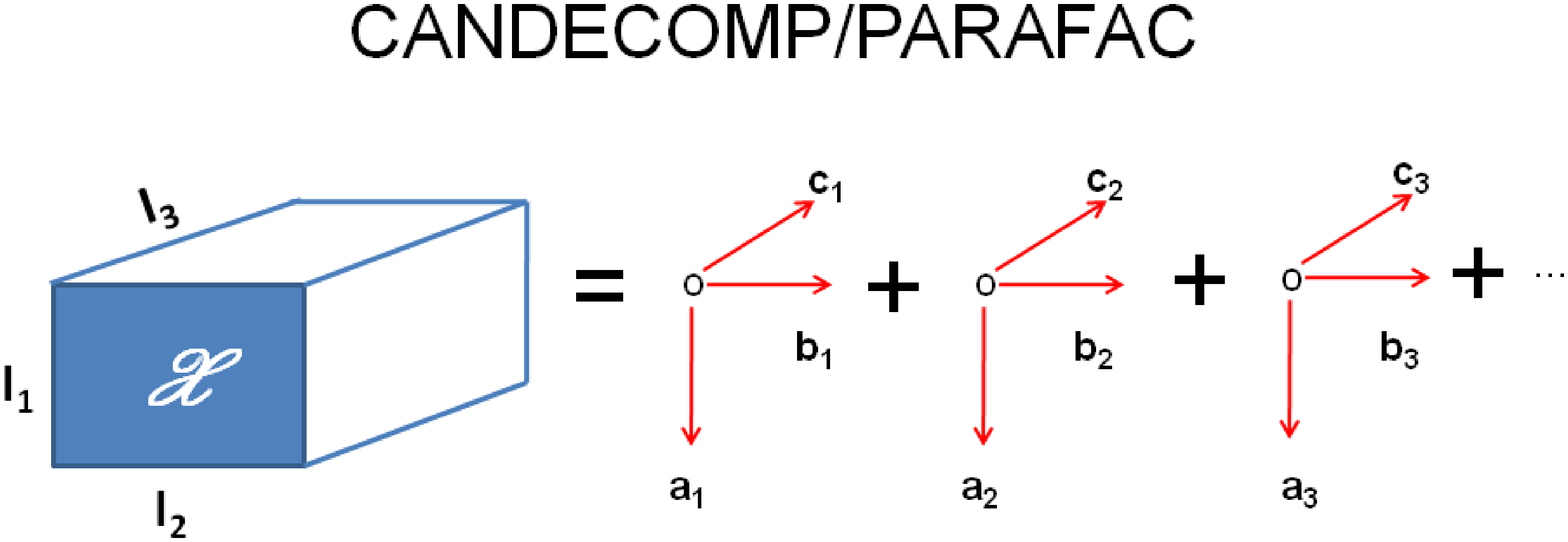,width=0.45\textwidth} 
\end{tabular}
\label{fig:fig2}
\caption{CANDECOMP/PARAFAC and Tucker tensor decompositions.  }
\end{figure}

\paragraph{Historical Remarks}  Tensors traditionally have been used in physics (e.g., stress and strain tensors). 
After Einstein presented the theory of general relativity tensor analysis became popular.
Certain ideas on multi-way analysis data back in 1944 and 1952 and are due to Raymond Cattell \cite{cat1,cat2}.
Tucker introduced tensor analysis in psychometrics \cite{tucker3} (Tucker family). 
Harshman \cite{harshman} and Carrol and Chang \cite{carroll} independently
proposed the canonical decomposition of a tensor (CANDECOMP family). 
These two families of decompositions come with different names, see \cite{tamarasurvey}.
The difference between them is visualized for a three way tensor in figure~\ref{fig:fig2}.
In the following we will focus on Tucker decompositions. 

\paragraph{Tensor Concepts}

Let ${\cal X} \in \field{R}^{I_1 \times I_2 \times \ldots \times I_d}$ be a multiway array.
We will call ${\cal X}$ a tensor, i.e., we will use the terms multiway array and tensor
interchangeably. The order of a tensor is the number of dimensions, also known as ways, modes
or aspects and is equal to $d$ for tensor ${\cal X}$. The dimensionality of the $j$-th mode is
equal to $I_j$. 

The norm of tensor ${\cal X}$ is defined to be the square root of the sum of all entries of the tensor
squared, i.e.,
\begin{equation}
|| {\cal X}||= \sqrt{ \sum_{j_1=1}^{I_1} \sum_{j_2=1}^{I_2} \ldots \sum_{j_d=1}^{I_d} x_{j_1,\ldots,j_d}^2}
\label{eq:norm}
\end{equation}
As we see the norm of a tensor is the straight-forward generalization of the Frobenius norm of a matrix
(2 modes) to $N$ modes.

The inner product of two tensors with the same number of modes and equal dimensionality per mode,
${\cal X},{\cal Y} \in \field{R}^{I_1 \times I_2 \times \ldots \times I_d}$, is defined by the following
equation:
\begin{equation}
\left\langle X,Y \right\rangle = \sum_{j_1=1}^{I_1} \sum_{j_2=1}^{I_2} \ldots \sum_{j_d=1}^{I_d} x_{j_1,\ldots,j_d} y_{j_1,\ldots,j_d}
\end{equation}
Observe that  equation~\ref{eq:norm} can equivalently be written as $|| {\cal X}||= \sqrt{\left\langle X,X \right\rangle}$
A tensor fiber (slice) is a one (two)-dimensional fragment of a tensor, obtained by fixing all indices but one (two).
For more details on tensor fibers and slices, see \cite{tamarasurvey}. 

\begin{figure}[h]
\begin{tabular}{c}
\psfig{figure=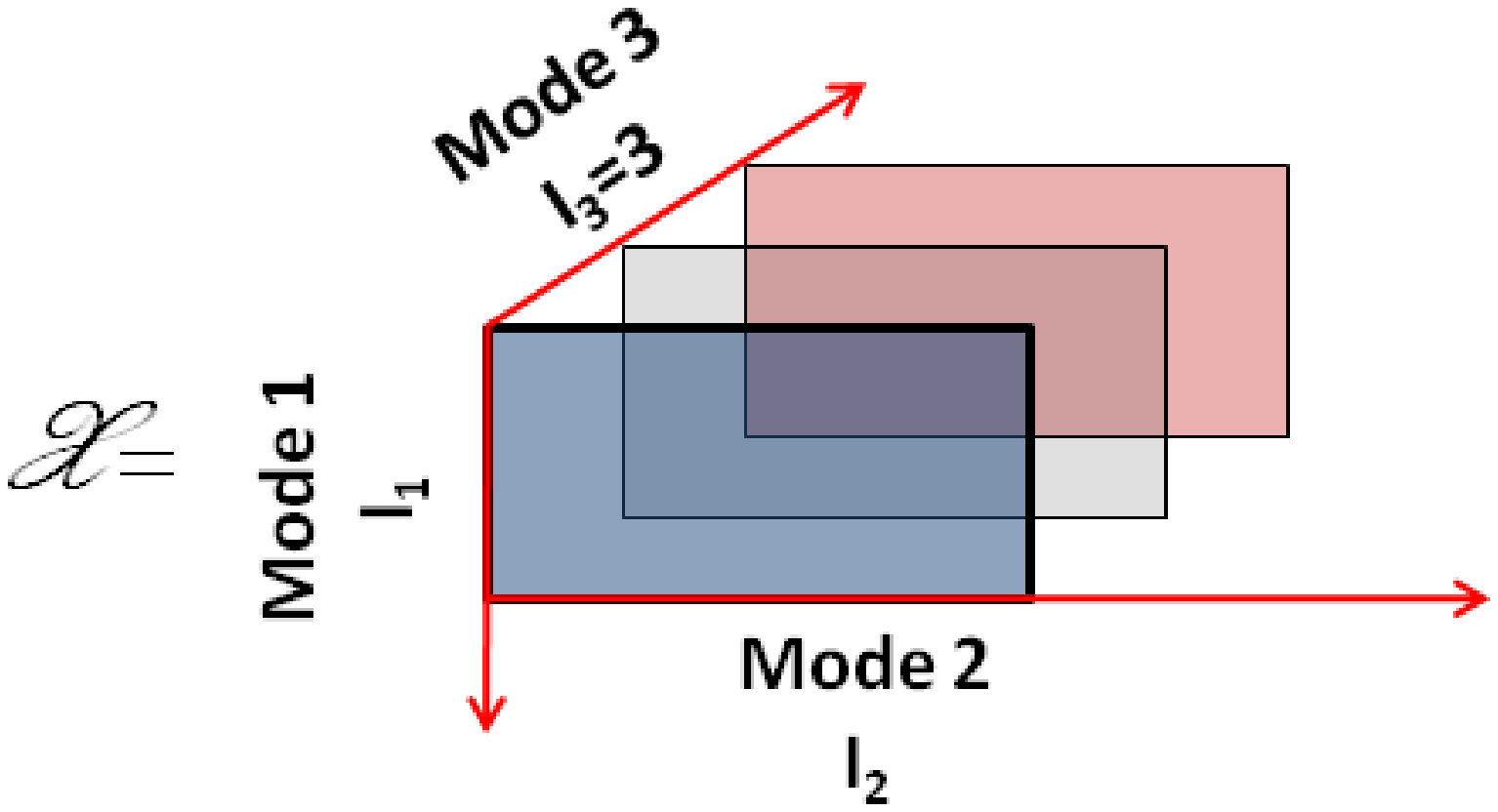,width=0.45\textwidth} \\
\psfig{figure=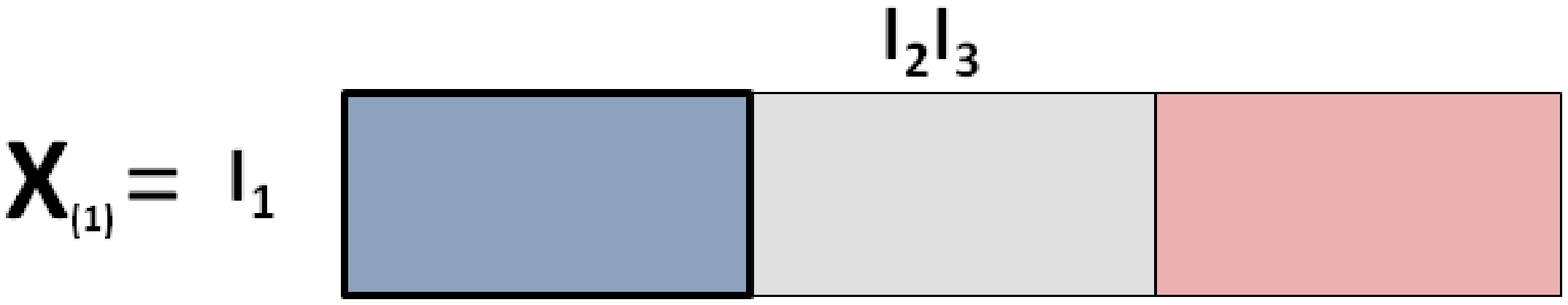,width=0.45\textwidth} 
\end{tabular}
\label{fig:fig3}
\caption{Matricization of a three-way $I_1 \times I_2 \times I_3$, $I_3=3$, tensor along the first mode. The three 
slices are denoted with different color.}
\end{figure}

Matricization along mode $k$, results in a $I_k  \times \prod_{j=1, j \neq k}^d I_j$ matrix. The $(i_1,\ldots,i_d)$ element
is mapped to $(i_k,j)$ where $j=1 + \sum_{q=1, q \neq k}^d (i_q-1) J_q$ where $J_q= \prod_{m=1, m \neq k}^{q-1} I_m$. 
Figure~\ref{fig:fig3} shows the concept of matricization for a three-way tensor. 
The operation of matricization naturally introduces the concept of a  vector containing ranks $(r_1,\ldots,r_d)$: $r_i$ is equal
to the rank of the $X_{(i)}$, the matrix resulting by the matricization of the tensor ${\cal X}$ along the $i$-th mode.

The $n$-mode product of ${\cal X}$ with a matrix $M \in \field{R}^{J \times I_n}$ is denoted by ${\cal X} \times_n M$
and is a tensor of size $I_1 \times I_2 \times \ldots I_{n-1} \times J \times I_{n+1} \times \ldots I_d$.
Specifically,
\begin{equation}
({\cal X} \times_n M )_{i_1 \ldots i_{n-1} j i_{n+1} \ldots i_d} = \sum_{i_n=1}^{I_n} x_{i_1 i_2 \ldots i_d} m_{ji_n}
\end{equation}
Some important facts concerning $n$-mode products, is the following: 

\begin{equation}
{\cal X} \times_m A \times_n B = {\cal X} \times_n B \times_m A, m\neq n
\end{equation}

The importance of this equation lies in the fact that the order of execution
of the tensor matrix products does not play any role, as long as the multiplications
are along different modes. 
When we multiply a tensor and two matrices along the same mode the following equation holds:

\begin{equation}
{\cal X} \times_m A \times_m B = {\cal X} \times_m (BA)
\label{eqb}
\end{equation}

Furthermore, if $UU^T=I$ then the following equation holds: 
\begin{equation}
|| A \times_n U ||= ||A||
\label{eq:eqa}
\end{equation}

The rank $R$ of the $d$-way tensor ${\cal X}$ is the minimum number of $d$-linear components
to fit ${\cal X}$ exactly, i.e.,: 
\begin{equation}
{\cal X} = \sum_{m=1}^R c_m^{(1)} \circ c_m^{(2)} \circ \ldots \circ c_m^{(d)} 
\end{equation}

where $c_1^{(j)},\ldots,c_R^{(j)}$ are the $R$ components for the $j$-th mode and 
$\circ$ denotes the tensor product. 
Even if the above generalization is a straightforward generalization of the rank of a matrix,
the concept of the tensor rank is special.  For example,  
for a matrix $A \in \field{R}^{2 \times 2}$ the column rank $R_c$ and the row rank $R_r$
are equal $R_c=R_r=r$ to the matrix rank $r$. Furthermore, $r \leq 2$. 
However for a tensor ${\cal X} \in \field{R}^{2 \times 2 \times 2}$ the rank can be 2 or 3 \cite{120567}.
Therefore the word rank can have different meanings:
a) The individual rank, i.e., for a specific instance of a tensor what is $R$?
b) The typical rank is the rank that we almost surely observe. For example for $2 \times 2 \times 2$
tensors the typical rank is $\{ 2,3\}$. c) Vector of ranks $(r_1,\ldots,r_d)$. The value of $r_i$
is equal to the rank of the matricized version $X_{(i)}$ of the tensor. 

Consider figure~\ref{fig:fig2}, which depicts a three mode tensor ${\cal X} \in \field{R}^{I_1 \times I_2 \times I_3}$. 
The PARAFAC/CANDECOMP model is given by equation~\ref{eq:parafac}, whereas the Tucker model is given by equation~\ref{eq:tucker}.

\begin{equation}
{\cal X}_{ijk} = \sum_{r=1}^R \alpha_{ir} b_{jr} c_{qr} \lambda_r + e_{ijk}
\label{eq:parafac}
\end{equation}

\begin{equation}
{\cal X}_{ijk} = \sum_{p=1}^P \sum_{q=1}^Q \sum_{r=1}^R \alpha_{ip} b_{jq} c_{kr} g_{pqr} + e_{ijk}
\label{eq:tucker}
\end{equation}

Few brief remarks on the above two models: 
a) In terms of the fit, the Tucker family is at least as good as the PARAFAC/CANDECOMP
since as we see from the above equations, the PARAFAC model can be viewed as a restrictive Tucker model, where the core tensor ${\cal G}$ 
is superdiagonal, i.e., $g_{pqr} \neq 0$ only if $p=q=r$. 
However, it is worth noting that better fit is not necessarily optimal (see \cite{brobook}, Ch.7)
b) The Tucker model does not result in unique solutions since it has rotational freedom. 
Typically one chooses a solution that satisfies a certain criterion, as the all-orthogonality
core tensor: $\left\langle G(m,:,:),G(n,:,:) \right\rangle = \left\langle G(:,m,:),G(:,n,:) \right\rangle= \left\langle G(:,:,m),G(:,:,n) \right\rangle=0$ when $m\neq n$ (\cite{citeulike:4308452}). 
c) Basic concepts as the uniqueness of the canonical tensor decomposition, degeneracy of the rank, border rank are not discussed. A good reference is \cite{tamarasurvey}
and the related references therein. 

In the following we focus on the Tucker family.
Compressing $n$ out of the $d$ modes of a tensor results in a Tucker-$n$ decomposition (\cite{kiers2000}). 
For example, for a three mode tensor we can have the Tucker1, Tucker2 and Tucker3 decomposition. 
In the following we discuss algorithms for the Tucker3 decomposition and briefly state some facts about Tucker2 and Tucker1
decompositions. Generalization to $d$ modes is straightforward. 

\paragraph{Tucker3 Algorithms} 

The algorithm which should be used to compute the Tucker3 decomposition of a tensor 
depends on whether or not the data is noise free. In the former case, an exact, closed
form solution exists, whereas in the latter case the alternating least squares algorithm (ALS)
is frequently used. However, it is worth noting that even in cases where there is noise in the data, 
the closed form solution a.k.a. as HOSVD \cite{tamarasurvey,citeulike:4308452} is satisfactory in practice \cite{luo-2009}. 

Let ${\cal X} \in \field{R}^{I_1 \times I_2 \times I_3}$ and $(r_1,r_2,r_3)$ the vector containing 
the desired approximation ranks along each mode. 
In the case of noise-free data, the algorithm matricizes the tensor along each mode
and computes the $r_k$ top left singular vectors $k=1,2,3$. Let $A_k$ be the $I_k \times r_k$ matrix
containing in its columns those vectors. The core tensor is computed with the following equation:
\begin{equation}
{\cal G} = {\cal X} \times_1 A_1^T \times_2 A_2^T \times_3 A_3^T
\label{eq:eqcore}
\end{equation}

In the case of noise in the data, one performs the alternating least squares algorithm. 
To solve the nonlinear optimization problem that tries to optimize the fit of the low rank approximation
with respect to the original tensor, one converts the problem into a linear one, by ``fixing'' all modes
but one and optimizing along that mode. This method is also known as Higher Order Orthogonal Iteration (HOOI). 
This procedure is continued until some stopping criterion is met, i.e., $\epsilon$ improvement in terms of fit.

\paragraph{Further Remarks}
H{\aa}stad proved that the tensor rank is an NP-complete problem \cite{DBLP:journals/jal/Hastad90}.
Lek-Heng Lim has proposed a theory for eigenvalues, eigenvectors, singular values  and singular vectors
\cite{DBLP:journals/corr/abs-math-0607648}. Maximum constraint satisfaction problems  (MAX-rCSP)
have been casted as a tensor decomposition problem (sum of rank one components).
In \cite{1060701} is proved that there is a PTAS (polynomial time approximation scheme) for a family 
of MAX-rCSP (i.e., core-dense). Sheehan and Saad in \cite{saad} give a unified view of different
dimensionality reduction techniques under the tensor framework.
A wealth of applications that use tensor decompositions exist, \cite{tamarasurvey} contains a wealth of such references.

\subsection{SVD and Fast Low Rank Approximation} 

Any matrix  $A \in \field{R}^{m \times n}$ can be written as a sum of rank one matrices, i.e.,
$A = \sum_{i=1}^r \sigma_i u_i v_i^T$,
where $u_i, i = 1 \ldots r$ (left singular vectors) and $v_i, i = 1 \ldots r$ (right singular vectors) 
are orthonormal and the singular values are ordered in decreasing order
$\sigma_1 \geq \ldots \geq \sigma_r > 0$. Here $r$ is the rank of $A$. 
We denote with $A_k$ the $k$-rank approximation of $A$, i.e., $A_k = \sum_{i=1}^k \sigma_i u_i v_i^T$. 
Among all matrices $C \in \field{R}^{m \times n}$ of rank at most $k$, 
$A_k$ is the one that minimizes $||A-C||_F $ (\cite{Horn:1985:MA}).
Since the computational cost of the SVD is high, $O( \min{(m^2n, n^2m)})$ for the full SVD 
approximation algorithms that give a close to the optimal solution $A_k$ have been developed. 
Frieze, Kannan and Vempala showed in a breakthrough paper \cite{1039494} that an approximate SVD 
can be computed by a randomly chosen submatrix of $A$. It is remarkable that the complexity
does not depend at all on $m,n$. Their Monte-Carlo algorithm with probability at least $1-\delta$ 
outputs a matrix $\hat{A}$ of rank at most $k$ 
that satisfies the following equation: 
\begin{equation} 
||A-\hat{A}||_F^2 \leq ||A-A_k||_F^2 + \epsilon ||A||_F^2
\label{eq:frieze}
\end{equation}
Drineas et al. in \cite{petros} showed how to find such a low rank approximation in $O(mk^2)$ time. 
A lot of work has followed on this problem. Here, we present the results of Achlioptas-McSherry \cite{Achlioptas01fastcomputation} 
which are used in our work\footnote{We call our proposed method MACH, to acknowledge the fact that 
it is based on the \textbf{Ach}lioptas-\textbf{M}cSherry work. }. The main theorem that is of interest to us is theorem~\ref{thm:ach}.

\begin{theorem}[Achlioptas-McSherry \cite{Achlioptas01fastcomputation}]
Let $A$ be any $m\times n$ matrix where $ 76 \leq m \leq n$ and let $b= \max_{ij}|A_{ij}|$.
For $p \geq (8 \log{n})^4/n$.
Let $\hat{A}$ be a random $m \times n$ matrix whose entries are independently distributed, with 
$\hat{A}_{ij} = A_{ij}/p$ with probability $p$ and 0 with probability $1-p$. 
Then with probability at least 1-exp$(19(\log n)^4)$, the matrix $N=A-\hat{A}$ satisfies the following two equations:

\begin{equation} 
||N_k||_2 < 4b\sqrt{\frac{n}{p}} 
\label{eq:2normach} 
\end{equation} 

\begin{equation} 
||N_k||_F < 4b\sqrt{\frac{nk}{p}} 
\label{eq:eqfrob} 
\end{equation}

\label{thm:ach}
\end{theorem}

\paragraph{Randomized Tensor Algorithms}

As already discussed, the most computationally expensive step for the Tucker decomposition 
is the SVD part. To alleviate this cost, two randomized algorithms which select columns according to a biased probability 
distribution for tensor decompositions \cite{Drineas05arandomized} have been proposed, extending the results of \cite{1109681}and \cite{Drineas04fastmonte}
to the multiway case and TensorCUR \cite{DBLP:conf/kdd/MahoneyMD06}, the extension of the CUR method \cite{md-cmdid-2009} in $n$-modes.
Roughly speaking, the bounds proved are of the form~\ref{eq:frieze}.
Another approach to approximating the Tucker decomposition for the case of a three-way tensor is presented in \cite{1461977}.
The proposed method matricizes the tensor as in all aforementioned algorithms and employes appropriately the 
matrix approximation described in \cite{goreinov}.

\section{Proposed Method}
\label{sec:meth}
\begin{figure*}[htb] 
\begin{center} 
\begin{tabular}{ccc}
\psfig{figure=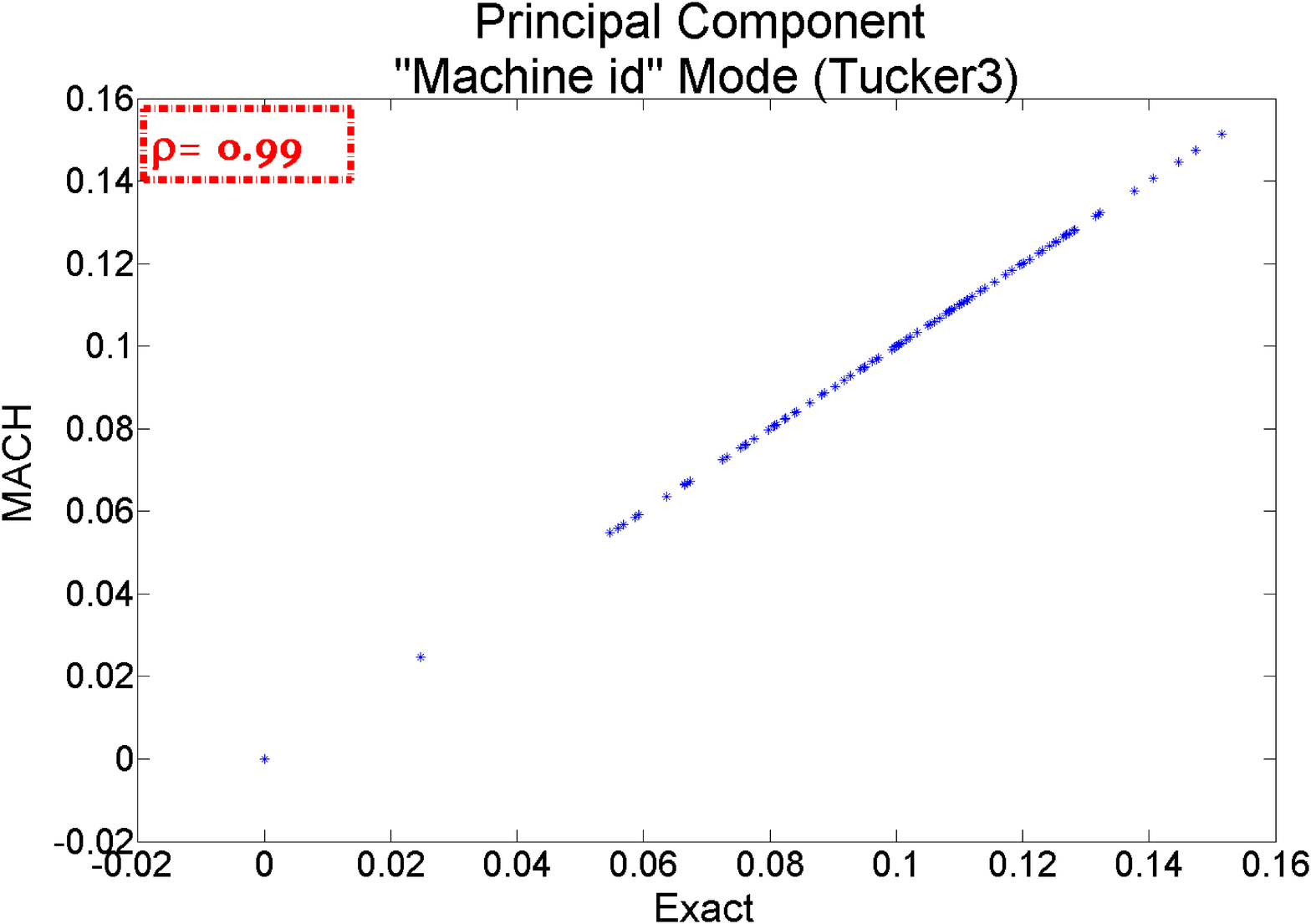,width=0.35\textwidth} & 
\psfig{figure=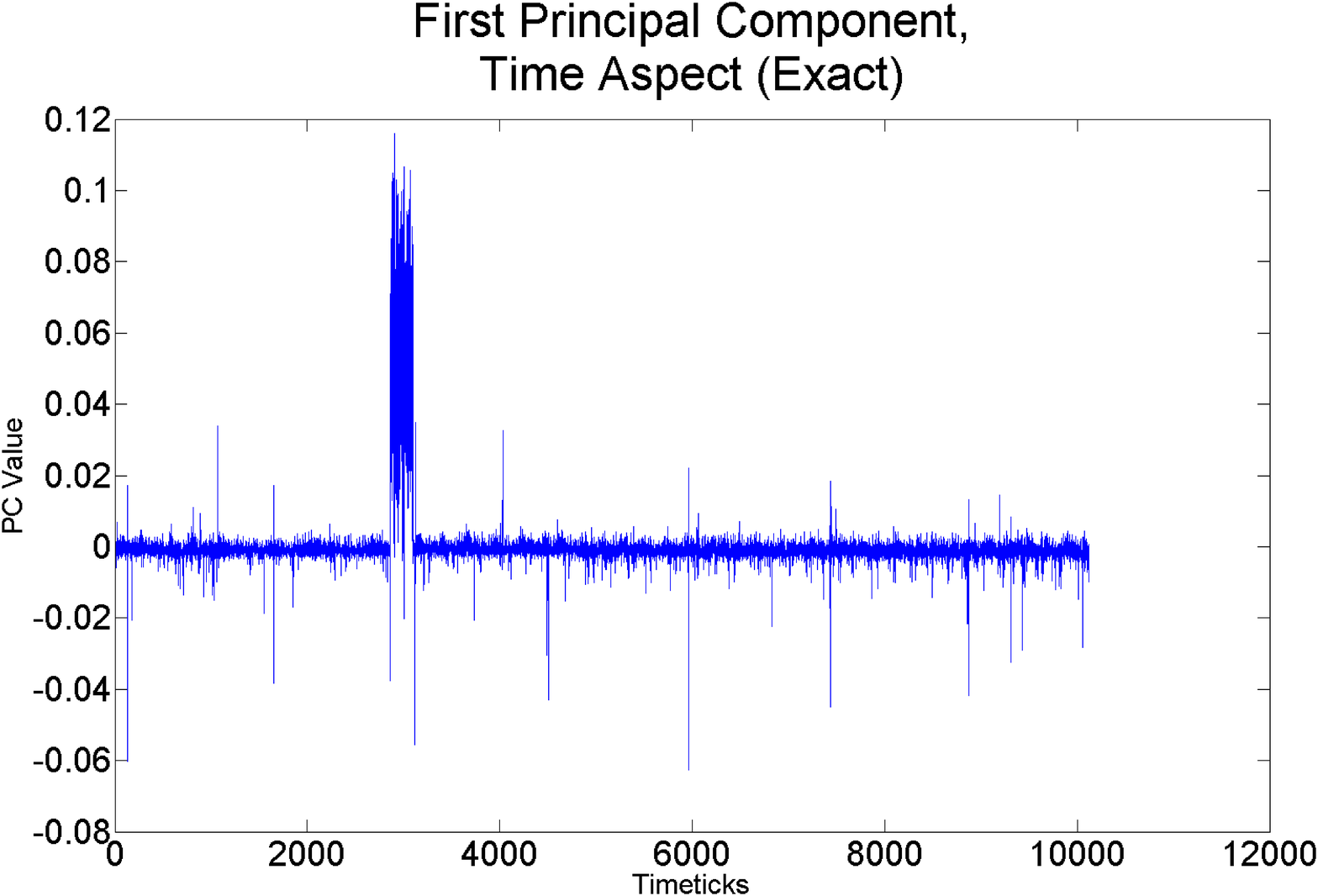,width=0.35\textwidth} & 
\psfig{figure=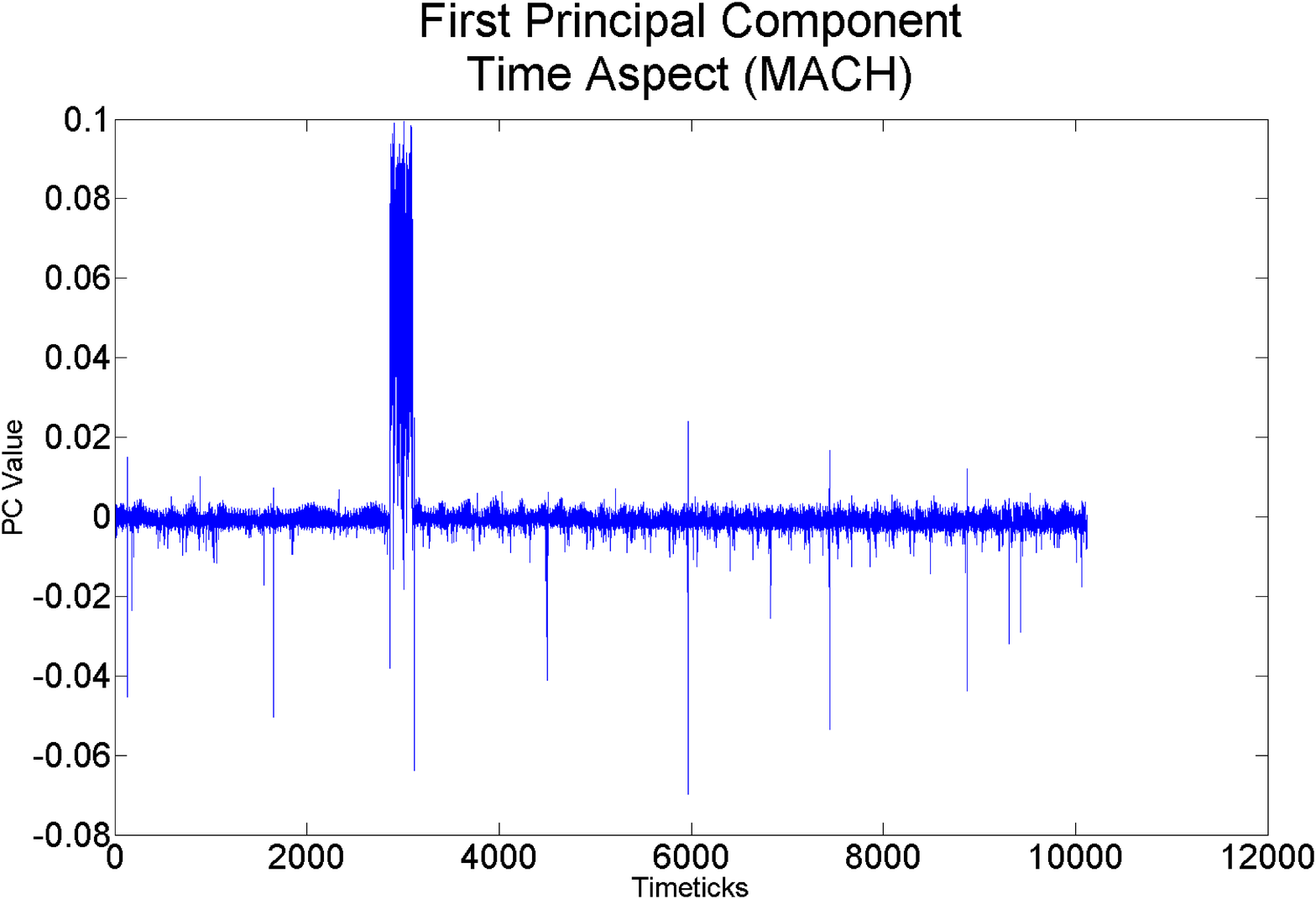,width=0.35\textwidth} \\
(a) & (b) & (c)
\end{tabular}
\caption{ (a) Top approximate Principal Component (PC) of the ``machine-id'' mode using the sampling MACH method vs. the exact PC. 
The PC was computed using a Tucker3 decomposition of the three-way tensor  machine id \textit{x} type of measurement \textit{x} timeticks,
formulated by data from the CMU monitoring system \cite{intemon}.
MACH used approximately 10\% of the original data. Pearson's correlation coefficient
is shown in the inset, and is almost equal to the ideal
value 1. Such PCs are of high practical value since they
are used in outlier detection algorithms \cite{intemon,1083674,citeulike:3637284}.
(b) Exact PC for the time aspect 
(c) Approximate PC using MACH. Pearson's correlation coefficient for the two time series equals 0.9772, again close to 
the ideal value 1.  }
\label{fig:intemonpctime}
\end{center}
\end{figure*}

\begin{figure*}[htb] 
\begin{center} 
\begin{tabular}{cc}
\psfig{figure=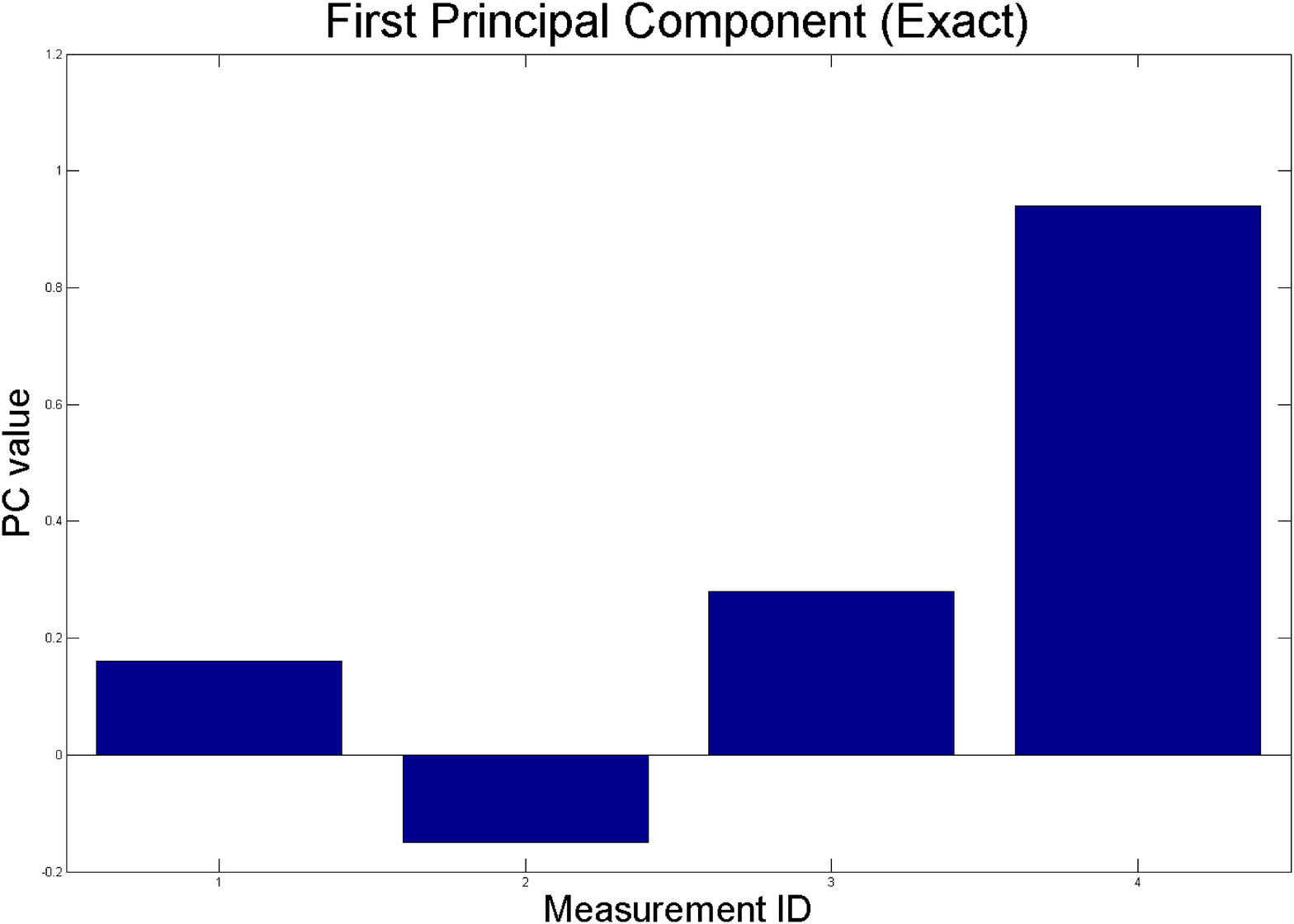,width=0.45\textwidth} & 
\psfig{figure=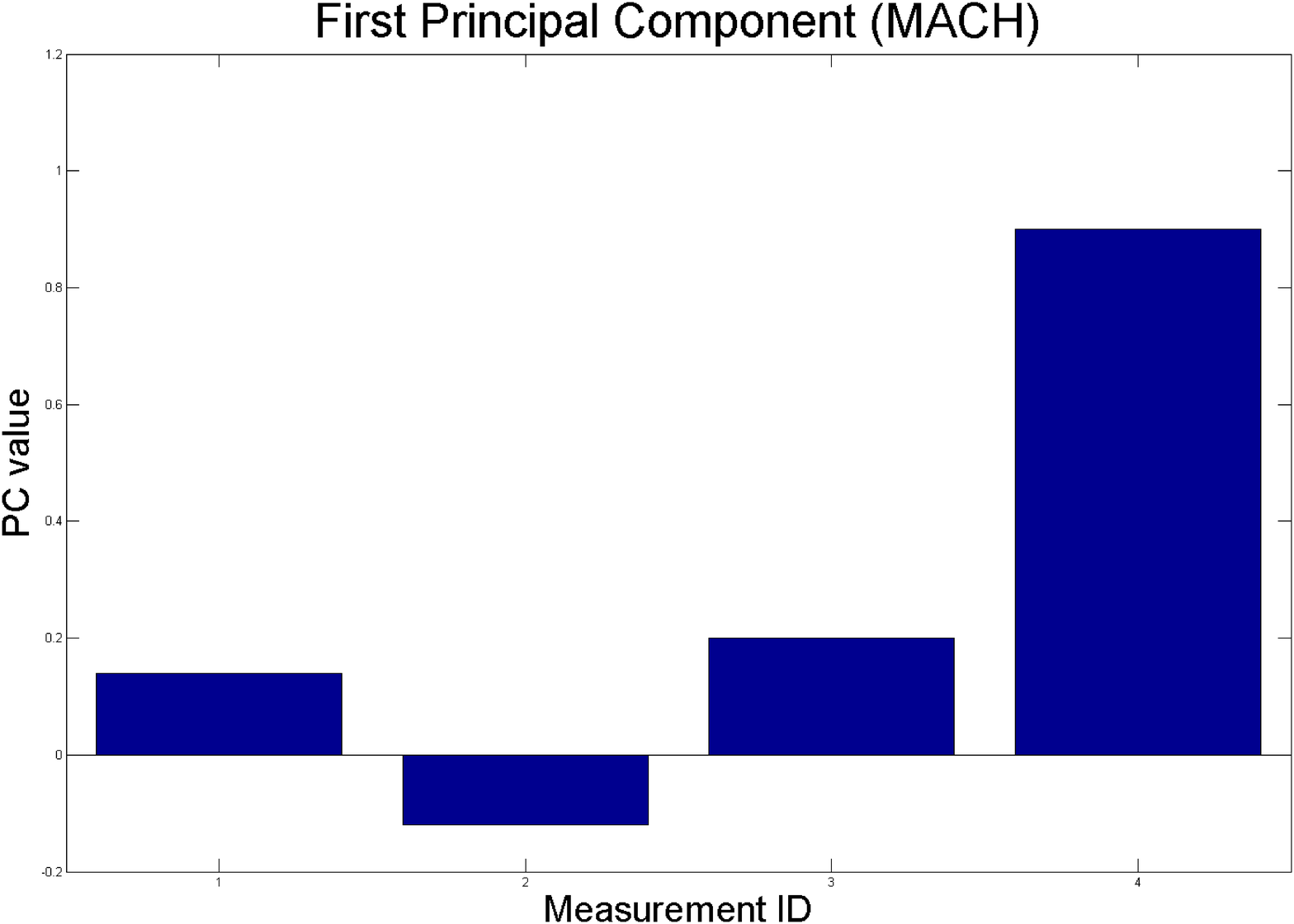,width=0.45\textwidth} 
\end{tabular}
\caption{ Principal component for the ``type of measurement'' aspect for the Intel Lab Berkeley sensor network \cite{1316741}.
Ids 1 to 4 correspond to voltage, humidity, temperature and light intensity. As we observe, the PC captures the correlations
between those types and MACH succeeds with p=0.1 in preserving them accurately.}
\label{fig:period2}
\end{center}
\end{figure*} 

The proof of theorem \ref{thrm:thrm1} follows: 

\begin{proof}

Let ${\cal E} = {\cal X} - \breve{{\cal X}}$ where $\breve{{\cal X}} = {\cal \hat{X}} \times_1 A^{(1)}A^{(1)T}  \ldots \times_d A^{(d)}A^{(d)T}$. 
Without loss of generality, let's assume $t$ of equation~\ref{eq:eqt} is minimum for index $d$, the last mode. 
Observe first that matrix $P_i=A^{(i)}A^{(i)T}$ for $i=1,\ldots,d$ is an orthogonal projector.
Specifically, $P_i$ projects on the subspace spanned by the top $r_i$ left singular vectors 
of the $i$-th matricization of tensor $\cal{\hat{X}}$. 
Therefore we have the following: 

$
|| {\cal E} || = ||{\cal X} -  {\cal \hat{X}} \times_1 A^{(1)}A^{(1)T} \times_2  \ldots \times_d A^{(d)}A^{(d)T}  || =
||{\cal X} - {\cal \hat{X}} \times_d A^{(d)}A^{(d)T} + {\cal \hat{X}} \times_d A^{(d)}A^{(d)T} -{\cal \hat{X}} \times_1 A^{(1)}A^{(1)T}  \ldots \times_d A^{(d)}A^{(d)T}  || \leq
||{\cal X} - {\cal \hat{X}} \times_d A^{(d)}A^{(d)T}|| + || ({\cal \hat{X}} -{\cal \hat{X}} \times_1 A^{(1)}A^{(1)T}  \ldots \times_{d-1} A^{(d-1)}A^{(d-1)T} ) \times_d A^{(d)}A^{(d)T}  || \leq
||{\cal X} - {\cal \hat{X}} \times_d A^{(d)}A^{(d)T}|| + || {\cal \hat{X}} -{\cal \hat{X}} \times_1 A^{(1)}A^{(1)T}  \ldots \times_{d-1} A^{(d-1)}A^{(d-1)T} ||
$

We obtained the above inequality by  adding and subtracting tensor ${\cal \hat{X}} \times_d A^{(d)}A^{(d)T}$ and applying 
the triangle inequality for a norm. The last line was obtained by using the fact that $ A^{(d)}A^{(d)T} $ 
is a projector thus we can only reduce the norm if we project the $d$-th matricization 
of tensor $ {\cal \hat{X}} -{\cal \hat{X}} \times_1 A^{(1)}A^{(1)T}  \ldots \times_{d-1} A^{(d-1)}A^{(d-1)T} $ 
along the $d$-th mode. 

Now consider the term $||{\cal X} - {\cal \hat{X}} \times_d A^{(d)}A^{(d)T}||$. If we matricize this tensor along the $d$-th mode
the Frobenius norm remains unchanged. 
Therefore  $||{\cal X} - {\cal \hat{X}} \times_d A^{(d)}A^{(d)T}|| = ||X_{(d)}- \hat{X}_{(d),r_d}||$. 
Now we use the following inequality to further bound this residual norm: 
$||{\cal X} - {\cal \hat{X}} \times_d A^{(d)}A^{(d)T}|| = ||X_{(d)}- \hat{X}_{(d),r_d}|| 
\leq ||X_{(d)}- X_{(d),r_d}||  + 4b \sqrt{ \frac{r_d}{p} \prod_{k=1}^{d-1} I_k } +$
$ 4 (b  ||X_{(d),r_d}||)^{\frac{1}{2}} (\frac{r_d}{p} \prod_{k=1}^{d-1} I_k)^{\frac{1}{4}} $.

The last inequality follows by combining two arguments which appear in \cite{Achlioptas01fastcomputation}. 
Namely, for any matrices A and B, the following holds:
$|| A- B_k ||_F \leq || A-A_k||_F + 2 \sqrt{ ||(A-B)_k||_F||A_k||_F } + ||(A-B)_k||_F$
Now substituting for $A$ the matrix $X_{(d)}$ and for $B_k$  the matrix $\hat{X}_{(d),r_d}$ and using equation~\ref{eq:eqfrob} 
to upper-bound $||(A-B)_k||=||(X-\hat{X})_{r_d}||$
gives the last inequality, where $k=r_d$ in our case.
Observe that we can use equation~\ref{eq:eqfrob}  since the assumptions 
of Theorem~\ref{thm:ach} hold by our assumptions. 

Now consider the term $|| {\cal \hat{X}} -{\cal \hat{X}} \times_1 A^{(1)}A^{(1)T}  \ldots \times_{d-1} A^{(d-1)}A^{(d-1)T} ||$. 
We will recursively apply simple properties of a norm and of a projector. 
Specifically:

$|| {\cal \hat{X}} -{\cal \hat{X}} \times_1 A^{(1)}A^{(1)T}  \ldots \times_{d-1} A^{(d-1)}A^{(d-1)T} || = 
|| {\cal \hat{X}} - {\cal \hat{X}} \times_{d-1} A^{(d-1)}A^{(d-1)T} + {\cal \hat{X}} \times_{d-1} A^{(d-1)}A^{(d-1)T} -{\cal \hat{X}} \times_1 A^{(1)}A^{(1)T}  \ldots \times_{d-1} A^{(d-1)}A^{(d-1)T} || 
\leq || {\cal \hat{X}} - {\cal \hat{X}} \times_{d-1} A^{(d-1)}A^{(d-1)T}|| +|| ({\cal \hat{X}}-{\cal \hat{X}} \times_1 A^{(1)}A^{(1)T}  \ldots \times_{d-2} A^{(d-2)}A^{(d-2)T} ) \times_{d-1} A^{(d-1)}A^{(d-1)T}||
\leq || {\cal \hat{X}} - {\cal \hat{X}} \times_{d-1} A^{(d-1)}A^{(d-1)T}|| +  || {\cal \hat{X}}-{\cal \hat{X}} \times_1 A^{(1)}A^{(1)T}  \ldots \times_{d-2} A^{(d-2)}A^{(d-2)T}  ||.$

Again we used the triangle inequality plus the fact that we can only reduce the norm if we project. Now repeating the same procedure to the last term 
and observing that for term $k$ for k=1,..,d-1 the norm does not change if we matricize with respect to that mode, we obtain the following simple upper bound:

 $|| {\cal \hat{X}} -{\cal \hat{X}} \times_1 A^{(1)}A^{(1)T}  \ldots \times_{d-1} A^{(d-1)}A^{(d-1)T}||  \leq \sum_{k=1}^{d-1} || \hat{X}_{k} - \hat{X}_{k,r_k}||$

By combining the above results we get the desired inequality. 
Three final remarks: observe that $b$ is the maximum of any matricization of our tensor and 
it is clear that since the above procedure gives for each aspect $i$ an inequality of the form $|| X - \breve{{\cal X}}|| \leq t_i$ 
then $|| X - \breve{{\cal X}}|| \leq \min_i t_i$. Finally the probability of success follows as the product of 
the success probabilities along each mode $i$.

\end{proof}

\paragraph{Remarks}
\textbf{(1)} Theorem~\ref{thrm:thrm1} suggests algorithm 1, MACH-HOSVD.
The algorithm takes as input a tensor ${\cal X}\in \field{R}^{I_1 \times \ldots \times I_d}$
and a vector containing the desired ranks of approximation along each mode $(R_1,\ldots,R_d)$. 
MACH tosses a coin for each non-zero entry ${\cal X}_{i_1,\ldots,i_d}$ of the tensor with probability $p$ of keeping
it and $1-p$ for zeroing it. In case of keeping it,
we reweigh it, i.e., ${\cal X}_{i_1,\ldots,i_d} \leftarrow \frac{ {\cal X}_{i_1,\ldots,i_d}}{p}$.
Then we return as an approximation to the HOSVD of tensor ${\cal X}$ the HOSVD of tensor ${\cal \hat{X}}$. 
The key idea behind proposing this algorithm is that for any matricization along mode $k$ of tensor ${\cal X}$
we get that:

$||{\cal X} - {\cal \hat{X}} \times_k A^{(k)}A^{(k)T}|| = ||X_{(k)}- \hat{X}_{(k),r_k}|| 
\leq ||X_{(k)}- X_{(k),r_k}||  + 4b \sqrt{ \frac{r_k}{p} \prod_{m=1, m \neq k}^{d} I_m } +$
$ 4 (b  X_{(k),r_k})^{\frac{1}{2}} (\frac{r_k}{p} \prod_{m=1, m \neq k}^{d} I_m)^{\frac{1}{4}} $.

Intuitively if tensor ${\cal X}$ has a good $(r_1,\ldots,r_d)$ Tucker
approximation, then matricization  along mode $k$ has a good $r_k$ rank approximation. 
The sparsification allows us to approximate this low rank approximation  $X_{(k),r_k}$ by $\hat{X}_{(k),r_k}$.

\textbf{(2)}  Frequently small $r_i$'s result in a satisfactory approximation of the original tensor.
The sparsification process we propose due to its simplicity is easily parallelizable and can easily be adapted to the streaming case
\cite{intemon} by tossing a coin each time a new measurement arrives.
\textbf{(3)} Picking the optimal $p$ in a real world application can be hard, especially in the context we are interested in, 
i.e., monitoring systems, where data is constantly arriving. Another potential problem are the assumptions 
of the theorem which may be violated. Fortunately, this does not render MACH algorithm useless. On the contrary, 
picking a constant $p$ even for small tensors which do not satisfy the conditions of the theorem result turns out to be 
accurate enough to perform data analysis. Therefore, a practicioner in whose application 
constant factor speedups and space savings are significant can just choose a constant $p$. 
\textbf{(4)} The expected speedup depends on the ``under-the-hood'' method to find the top $k$ singular vectors of a matrix.
Lanczos method \cite{citeulike:2122238} is such a method. Recently, approximation algorithms approximate the $k$-rank approximation of a matrix
in linear time \cite{tamas}. Thus, if such a fast algorithm is used, the expected speedup is $\frac{1}{p}$. 
\textbf{(5)} Theorem~\ref{thrm:thrm1} refers to the HOSVD of a tensor. We can apply the same idea 
to the HOOI. This results in algorithm 2. We do not analyze the performance of algorithm 2 here, 
since it would require the analysis of the convergence of the alternating least squares method
which does not exist yet. As we will show in the experimental section~\ref{sec:exp}, MACH-HOOI gives 
satisfactory results.

\begin{algorithm}[!ht]
\caption{MACH-HOSVD} 
\begin{algorithmic}
\REQUIRE ${\cal X} \in \field{R}^{I_1 \times \ldots \times I_d}$
\REQUIRE $(r_1, \ldots, r_d)$
\REQUIRE $p$

\COMMENT{MACH}
\STATE \textbf{for} each ${\cal X}_{i_1,\ldots,i_d}$, $i_j=1\ldots I_j$ toss a coin with probability $p$ of keeping it.
\IF{ success }
\STATE ${\cal \hat{X}}_{i_1,\ldots,i_d} \leftarrow \frac{{\cal X}_{i_1,\ldots,i_d}}{p}$
\ELSE
\STATE ${\cal \hat{X}}_{i_1,\ldots,i_d} \leftarrow 0$
\ENDIF
\COMMENT{HOSVD}
\FOR{$i=1$ to $d$}
   \STATE $A^{(i)} \leftarrow r_i$ leading left singular vectors of $\hat{X}_{(i)}$ 
\ENDFOR
\STATE ${\cal G} \leftarrow {\cal \hat{X}} \times_1 A^{(1)T} \times_2 A^{(2)T} \ldots \times_d A^{(d)T}$
\RETURN ${\cal G}, A^{(1)}, \ldots, A^{(d)}$
\end{algorithmic}
\end{algorithm}

\begin{algorithm}[!ht]
\caption{MACH-HOOI} 
\begin{algorithmic}

\REQUIRE ${\cal X} \in \field{R}^{I_1 \times \ldots \times I_d}$
\REQUIRE $(r_1, \ldots, r_d)$
\REQUIRE $p$

\COMMENT{MACH}
\STATE \textbf{for} each ${\cal X}_{i_1,\ldots,i_d}$, $i_j=1\ldots I_j$ toss a coin with probability $p$ of keeping it.
\IF{ success }
\STATE ${\cal \hat{X}}_{i_1,\ldots,i_d} \leftarrow \frac{{\cal X}_{i_1,\ldots,i_d}}{p}$
\ELSE
\STATE ${\cal \hat{X}}_{i_1,\ldots,i_d} \leftarrow 0$
\ENDIF

\COMMENT{HOOI}	
\STATE initialize $A^{(k)} \in \field{R}^{I_k \times r_k}$ for $k=1\ldots d$ using HOSVD

\REPEAT
\FOR{$i=1$ to $d$}
   \STATE ${\cal Y} \leftarrow {\cal \hat{X}} \times_1 A^{(1)T} \ldots \times_{i-1}A^{(i-1)T} \times_{i+1}A^{(i+1)T} \ldots \times_{d}A^{(d)T} $
   \STATE $A^{(i)} \leftarrow r_i$ leading left singular vectors of $Y_{(i)}$ 
\ENDFOR
\UNTIL{fit stops improving or maximum number of iterations is reached} 
\STATE ${\cal G} \leftarrow {\cal \hat{X}} \times_1 A^{(1)T} \times_2 A^{(2)T} \ldots \times_d A^{(d)T}$
\RETURN ${\cal G}, A^{(1)}, \ldots, A^{(d)}$
\end{algorithmic}
\end{algorithm}

\section{Experiments}
\label{sec:exp}

\begin{figure}[h]
\begin{tabular}{c}
\psfig{figure=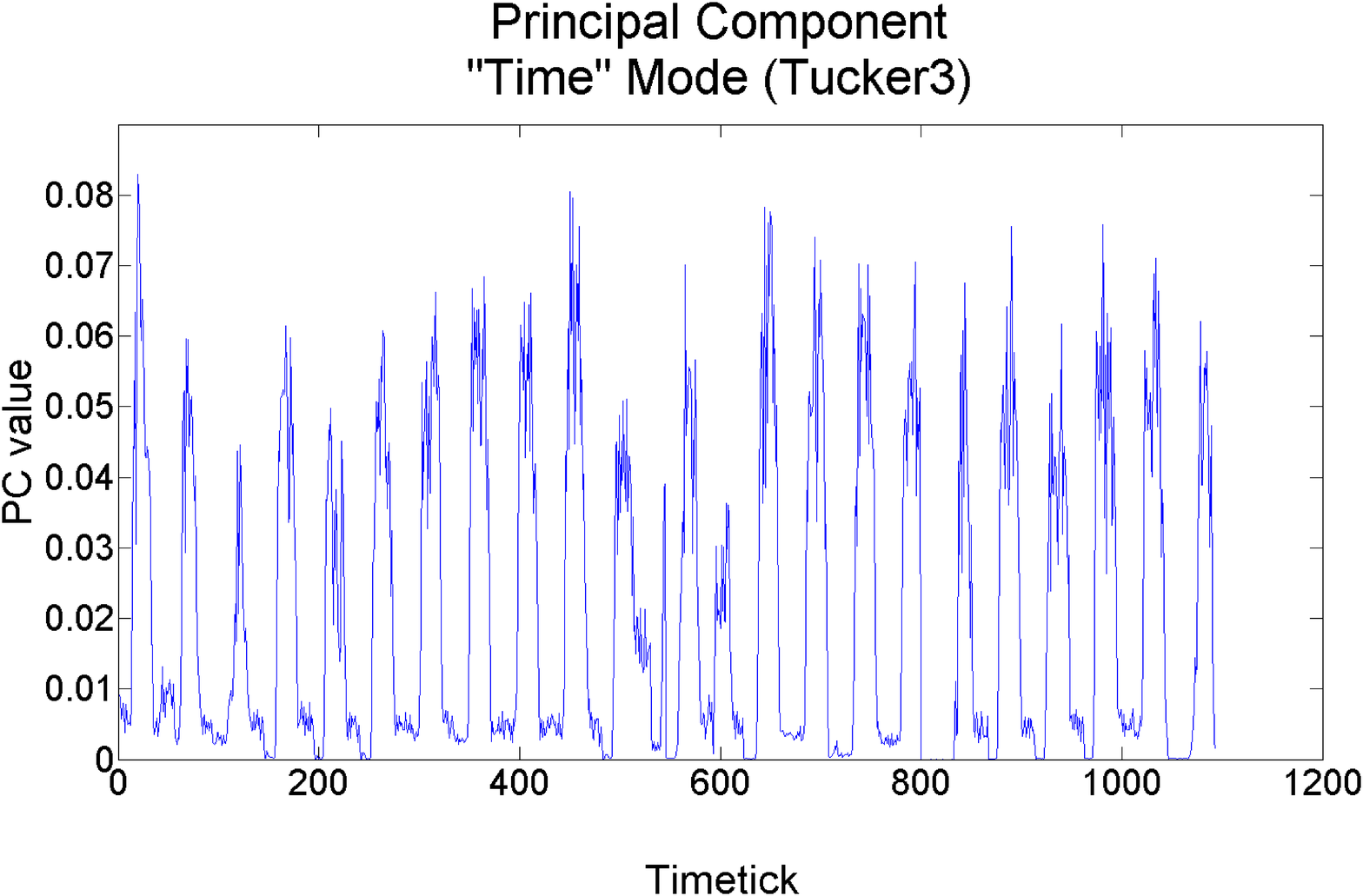,width=0.45\textwidth} 
\end{tabular}
\label{fig:aaa}
\caption{Principal component for the time aspect using MACH with p=0.1. Daily periodicity 
appears to be the dominant latent factor for the time aspect.}
\end{figure}

\paragraph{Experimental Setup} 

We used the Tensor Toolbox \cite{SAND2006-7592}, which contains 
MATLAB implementations of the HOSVD and the HOOI.
Our experiments ran in a 2GB RAM, Intel(R) Core(TM)2 Duo CPU at 2.4GHz Ubuntu Linux machine.
Table \ref{tab:datasets} describes the datasets we use. 
The motivation of our method as already mentioned, is to provide 
a practical algorithm for tensor decompositions which involve streams, such as monitoring systems.  
It is also worth noting that the assumptions of theorem~\ref{thrm:thrm1} do not hold. 
Nonetheless, results are close to ideal. 
Finally, in this section we report experimental results for the MACH-HOOI. The reason is that
Tucker decompositions using alternating least squares are used in practice more than the HOSVD 
and also, they have already been successfully applied to the real world problems we consider in the following \cite{citeulike:3637284}.
The results for HOSVD are consistently same or better than the results we report in this section. 

\begin{table}[htb] %
\begin{center}
  \begin{tabular}{|c|c|}
      \hline
    \textbf{name} & \textbf{$I_1 \times I_2 \times I_3$} 
    \\
    \hline
    {\tt Sensor } & 54-by-4-by 5385 \\
    {\tt Network Data (\cite{1316741})} &  \\ \hline
    {\tt Intemon  } & 100-by-12-by-10080 \\
    {\tt Data (\cite{intemon})} &  \\ 
    \hline
  \end{tabular}
\end{center}
 \caption{Dataset summary. The third aspect is the time aspect.}
\label{tab:datasets}
\end{table}

\begin{figure*}[htb] 
\begin{center}
  \begin{tabular}{cc}
    \resizebox{.33\hsize}{0.45\hsize}{ \psfig{figure=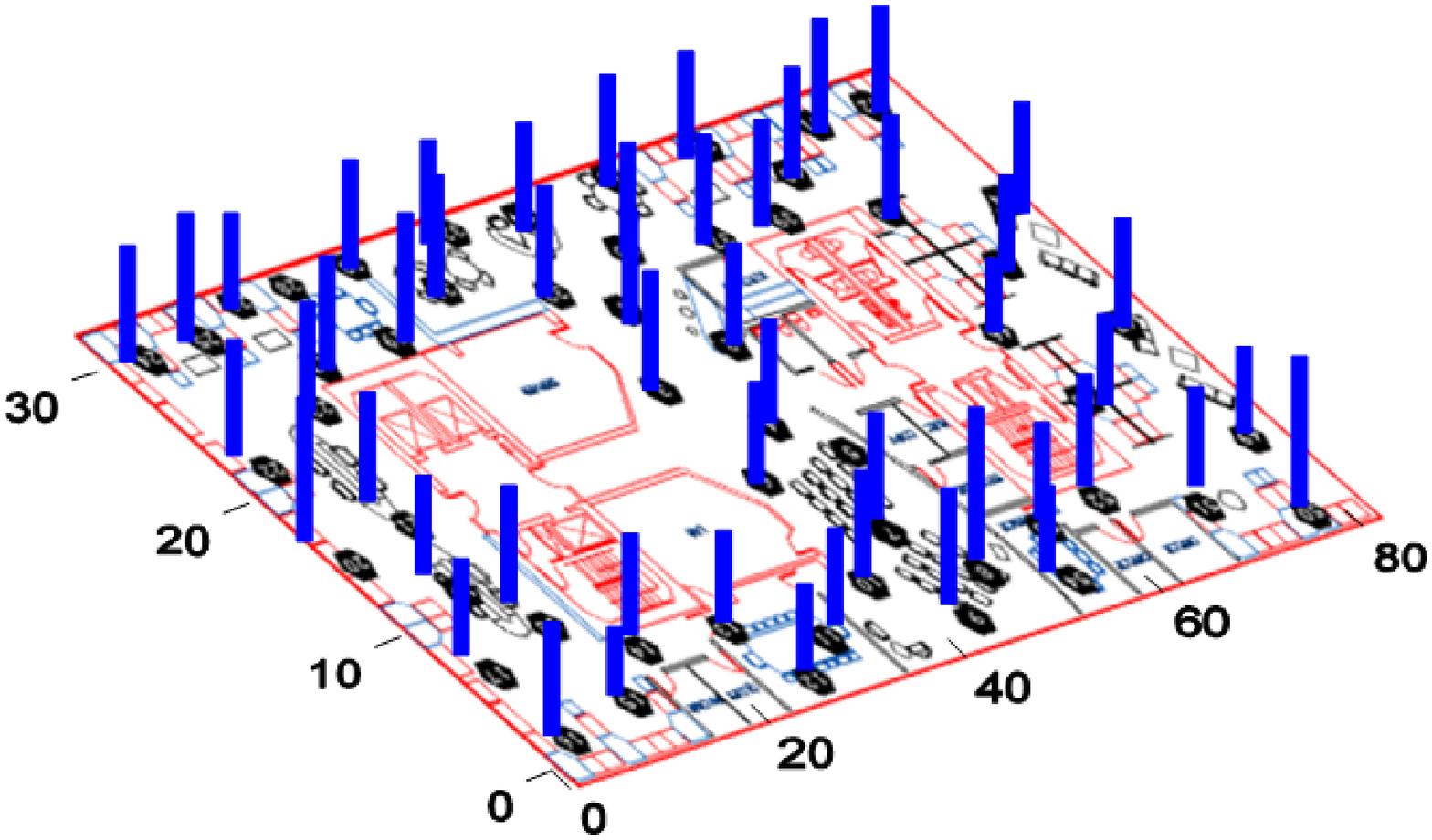}{0.9}}&
	\resizebox{.33\hsize}{0.45\hsize}{ \psfig{figure=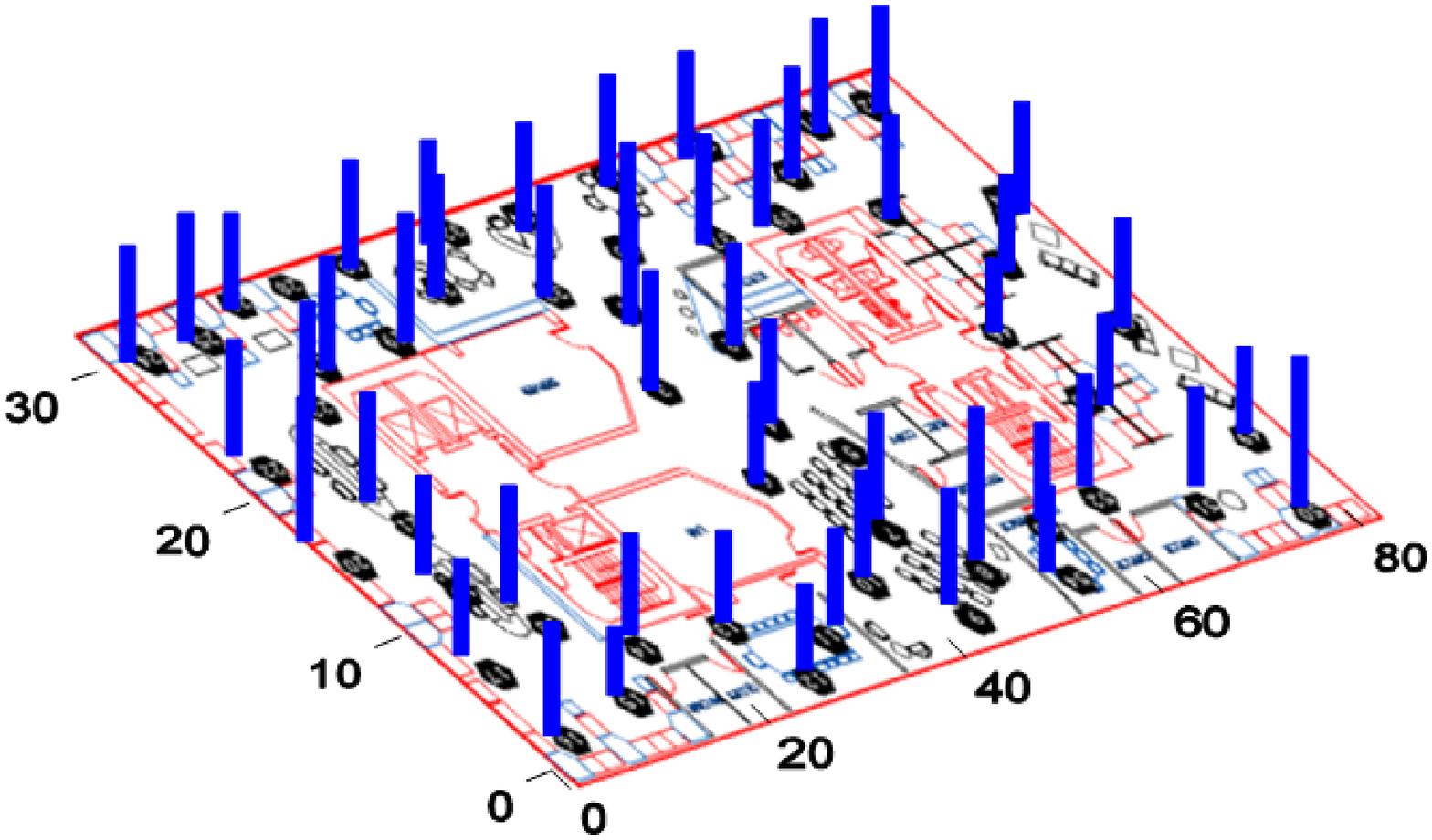}{0.9}} \\
 (a) SENSOR Concept 1 & (b) SENSOR Concept 1 using MACH
  \end{tabular}
\end{center}\vspace{-1\baselineskip} \caption{ 
(a) shows the distribution of the most dominant trend,
(b) shows the distribution of the most dominant trend, using MACH algorithm with p=0.1.
Pearson's correlation coefficient equals 0.93, and thus  the 
qualitative analysis of the dominant sensor/spatial correlations remains unaffected by the sparsification.
Colored bars indicate positive weights of the corresponding sensors. As suggested in \cite{citeulike:3637284},
e values assigned to the sensors are more or less uniform suggesting that 
the dominant trend is strongly affected by the daily periodicity. }
\label{fig:spatial-case}
\end{figure*}

\subsection{Monitoring computer networks}

As already mentioned in Section~\ref{sec:intro}, a prototype monitoring system in Carnegie Mellon University uses
data mining techniques successfully \cite{1083674,intemon,citeulike:3637284} to spot anomalies and detect
correlations among different types of measurements and machines. Analyzing and applying these techniques on
large amounts of data however is a challenge. A natural way to model this type of data is a three-way tensor, 
i.e., machine id$\times$type of measurement$\times$time. The data on which we apply MACH is a tensor 
${\cal X}\in \field{R}^{100 \times 12 \times 10080}$. The first aspect is the ``machine id'' aspect
and the second is the ``type of measurement'' aspect
(bytes received, unicast packets received, bytes sent, unicast packets sent, unprivileged CPU utilization,
other CPU utilization, privileged CPU utilization, CPU idle time, available memory, number of users, number of processes
and disk usage). The third aspect is the time aspect. 
Figure~\ref{fig:intemonpctime}(a) plots the Principal Component (PC) of the ``machine id'' aspect 
after performing a Tucker3 decomposition using MACH versus the exact PC.
Our sampling approach thus kept approximately the 10\% of the original data. 
As the figure shows, the results are close to ideal and similar results hold 
for the other few top PCs. Specifically, Pearson's correlation coefficient is 0.99, close to the ideal
1 which is the perfect linear correlation between the exact and the approximate top PC. 
This fact is important since these PCs can be used to find outlier machines, which ideally would be the machines that face a functionality problem.
Figures~\ref{fig:intemonpctime}(b),~\ref{fig:intemonpctime}(c) show the exact top and the MACH PC for the time aspect. Pearson's correlation
coefficient is equal to 0.98. We observe that there is no clear periodic pattern in this time series.
The important fact is that MACH using only 10\% of the data, results in a good approximation. 
This is of significant practical value and can be used also in conjunction with DTA
\cite{DTA} to perform dynamic tensor analysis in larger time windows. 

\subsection{Environmental Monitoring} 

In this application we use data from the Intel Berkeley Research Lab sensor network \cite{1316741}.
The data is collected from 54 Mica2Dot sensors which measure at every timetick
humidity, temperature, light and voltage. 

It has been shown in \cite{citeulike:3637284} that tensor decompositions along with a wavelet analysis can efficiently
capture anomalies in the network, i.e., battery outage as well as spatial and measurement correlations.
In this section we show that a random subset about 10\% of the initial data volume suffices to perform 
the same analysis as if we had used the whole tensor. 

Figure~\ref{fig:period2} shows the correlations revealed by the the principal component for
the ``type of measurement'' aspect. As we observe, voltage, temperature and light intensity 
are positively correlated, whereas at the same time the latter types of measurement are negatively correlated with
humidity.  This is because during the day, temperature and light intensity go up but humidity
drops because the air conditioning system is on. Similarly , during the night, temperature and light
intensity go down but humidity increases because the air conditioning system is off. Furthermore, 
the positive correlation between voltage and temperature is due to the design of MICA2 sensors.
As we observe again, MACH gives the same qualitative analysis by examining the principal component. 
Pearson's correlation coefficient is close to the ideal value 1. 
Figure 5 shows the principal component for the time aspect. A periodic pattern is apparent
and corresponds to the daily periodicity. Performing a Tucker2 decomposition as suggested by 
\cite{citeulike:3637284} and plotting the fiber of the core tensor corresponding to the principal components of the tensor 
for the  ``sensor id'' and ``measurement type'' mode, the results are again close to ideal.
Figure~\ref{fig:spatial-case}(a) shows the principal component for the ``sensor id''
aspect using the exact Tucker decomposition and Figure~\ref{fig:spatial-case}(b) using MACH with p=0.1.
The top component captures spatial correlations and MACH preserves them with a random subset of size approximately 10\% of the original data. 
Pearson's correlation coefficient is equal to 0.93.

\subsection{Discussion}

\paragraph{General}
The above experiments show MACH results in a good approximation of the desired low rank Tucker
approximation of a tensor. Similar result hold for the other few top principal components of the Tucker
decomposition. Also, as already mentioned, results for HOSVD are consistently better or same with the reported ones,
and the above applications were selected since 
it has already been shown by previous work that Tucker decompositions and SVD can 
detect anomalies and correlations. Thus, the main goal of this section is -rather than 
introducing new applications- to show that keeping a small
random subset of the tensor can give good results.

\paragraph{How to choose p?}
Choosing the best possible $p$ is an issue.
We use a constant p, i.e., p=10\%
in our experiments\footnote{For both applications that value of p, gives excellent results. If we set p=5\% for the first
application results get significantly worse whereas for the second results remain good.}. 
Constant $p$'s are of significant practical value in such settings where it is not clear how one should set $p$ to sparsify
the underlying tensor optimally. For ``post-mortem'' data analysis, one can try setting lower values for p according to 
theorem~\ref{thrm:thrm1}.

\paragraph{Speedups \& Synthetic Experiment} 
Speedups due to the small size of the two datasets and the implementation was less than the 
expected 10$\times$ (typically 2-3$\times$ faster performance).
However, as the size of the tensor grows bigger (i.e., the number of non-zeros) the speedup should become apparent.
For example consider a tensor ${\cal X} \in \field{R}^{n \times n \times n}$, with ${\cal X}_{ijk} = \frac{1}{i+j+k}$
and assume we want an $(r,r,r)$ approximation. 
As shown in \cite{1120100,Tyrtyshnikov} for an approximation with error $\epsilon$ the rank grows logarithmically with $n$ and $\epsilon$,
satisfying inequality~\ref{eq:ineq1}:
 
\begin{equation} 
r \leq C(\log{n} \log^2{\epsilon})
\label{eq:ineq1} 
\end{equation} 

This tensor appears in numerical solutions of integral equations \cite{1461977}. 
A small numerical example for $r=4$ and $n=200$ gives the results in table~\ref{tab:synthetic}
for $p=0.1$. The second column of the table contains a vector of three values $(\rho_1,\rho_2,\rho_3)$. 
$\rho_i$ i=1,2,3 is the correlation coefficient between the principal component of the exact 
Tucker3 decomposition and the MACH Tucker3 decomposition  of aspect i. 
As we see the correlation is almost perfect for all aspects. 
This is significant since the single important interaction
is betwen the first principal components. This can be seen by examining the core tensor\footnote{
The exact core tensor value which determines the interaction between the top PCs is $g(1,1,1)=$ 18.4856 and  18.4887 for 
the MACH decomposition. The next largest core tensor value has absolute value 2.61$<<$18.5.}
The third column contains the accuracy of the approximation, i.e., 1- $\frac{|| {\cal X} - {\cal \hat{X}} \times_1 A^{(1)}A^{(1)T} \times_2  \ldots \times_d A^{(d)}A^{(d)T} ||  }{||{\cal X}||}$. 
As we see the speedup now becomes apparent, i.e., 7.52$\times$ faster. Finally, when we attempt to run Tucker3 on a larger tensor
with n=500, MATLAB runs out of memory, whereas when using p=0.1 we can run the Tucker decomposition
and obtain an accurate precision. Observe that for this specific value of $n$ the assumptions of 
theorem~\ref{thrm:thrm1} do not hold, i.e., $\alpha = 200^2$ thus $p \geq \frac{(8 log{\alpha})^4}{\alpha}=\frac{(8 log{200^2})^4}{200^2} >>1 $. However results are satisfactory and this holds for even smaller values 
of $p$ as one can verify. 

\begin{table}[htb] %
\begin{center}
  \begin{tabular}{|c|c|c|}
      \hline \hline
    p   & $\rho$                 & accuracy \\ \hline
    0.1 & (0.9967, 0.9955, & 87\%     \\  
        &  0.9964)         &  \\ \hline \hline
    exact (sec) & MACH(sec) &  speedup ($\times$faster)  \\ \hline
    119.8     & 15.92     &  7.52 \\  \hline \hline
    
  \end{tabular}
\end{center}
 \caption{Synthetic experiment results }
\label{tab:synthetic}
\end{table}

\section{Conclusions}
\label{sec:concl}
In this paper we focused on Tucker decompositions. We proposed MACH, a simple randomized algorithm which
is embarassingly parallel and adapts easily to tensor streams, since it simply tosses
a coin for each entry of the tensor. Specifically, our contributions include:
\begin{itemize} 
   \item A new algorithm MACH, which keeps a small percentage of the entries of a tensor, 
         and still produces an accurate low rank approximation of the tensor. 
         We performed a theoretical analysis of the algorithm in Theorem~\ref{thrm:thrm1}
         and of its speedup in Section~\ref{sec:meth}.
   \item An experimental evaluation of MACH on two real world datasets, both generated from monitoring systems, 
          and on a synthetic one,  where we showed that for constant values of p excellent performance. 
\end{itemize} 

This algorithm will be incorporated in the PEGASUS software library \cite{hadi},a graph and tensor mining system for handling 
large amounts of data using Hadoop, the open source version of MapReduce \cite{citeulike:430834}. 
Given the effectiveness of the sampling approach and that it is embarassingly parallel, it will 
be useful when dealing with huge amounts of data, given of course that the empirical observation 
that low rank approximations are satisfactory in practice. 
The (in)effectiveness of MACH with respect to the PARAFAC/CANDECOMP decomposition will be investigated in future work.

\section{Acknowledgements}
The author would like to thank Gary L. Miller, Petros Drineas, Ioannis Koutis, M.N. Kolountzakis and Christos Boutsidis
for their feedback and Christos Faloutsos for his support and for introducing the author to tensor mining.

The author was supported in part by the National Science Foundation
under Grant No. IIS-0705359. Any opinions, findings, and
conclusions or recommendations expressed in this material
are those of the author(s) and do not necessarily reflect the
views of the National Science Foundation.

\bibliographystyle{abbrv}
\bibliography{mach}

\end{document}